\documentclass[showpacs,twocolumn,aps,prb,superscriptaddress]{revtex4}
\usepackage{amsmath}
\usepackage{amsfonts}
\usepackage{amssymb}
\usepackage{amsthm}
\usepackage{graphicx}

\begin{document}

\title{Spin transfer torque enhancement in dual spin valve in the ballistic
regime}
\author{P. Yan}
\affiliation{Physics Department, The Hong Kong University of Science and Technology,
Clear Water Bay, Hong Kong SAR, China}
\author{Z. Z. Sun}
\affiliation{Institute for Theoretical Physics, University of Regensburg, D-93040
Regensburg, Germany}
\author{X. R. Wang}
\affiliation{Physics Department, The Hong Kong University of Science and Technology,
Clear Water Bay, Hong Kong SAR, China}
\date{\today}

\begin{abstract}
The spin transfer torque in all-metal dual spin valve, in which two
antiparallelly aligned pinned ferromagnetic layers are on the two sides of a
free ferromagnetic layer with two thin nonmagnetic spacers in between, is
studied in the ballistic regime. It is argued that, similar to the results
in the diffusion regime, the spin transfer torque is dramatically enhanced
in comparison to that in a conventional spin valve although no spin
accumulation exists at the magnetic-nonmagnetic interfaces. Within the
Slonczewski's approach, an analytical expression of the torque on the free
magnetic layer is obtained, which may serve as a theoretical model for the
micromagnetic simulation of the spin dynamics in dual spin valve. Depending
on the orientation of free layer and the degree of electron polarization,
the spin transfer torque enhancement could be tens times. The general cases
when transmission and reflection probabilities of free layer are different
from zero or one are also numerically calculated.
\end{abstract}

\maketitle

\section{INTRODUCTION}

Current induced magnetization reversal of magnetic multilayers has attracted
much research interest due to its rich physics and potential applications in
spintronic devices. \cite{Katine,Grollier,Xia,Urazhdin1,Tsoi,Chen,Zhou,Su}
Spin valve, consisting of two ferromagnetic layers and one nonmagnetic
spacer in between, is one of such multilayer structures. In a spin valve,
one of the magnetic layers, acting as a spin polarizer, is thick so that
conducting electrons are polarized after passing through it. The polarized
conducting electrons will transfer their spin angular momentums to local
magnetization of the thinner free magnetic layer, resulting in the spin
transfer torque (STT) effect first proposed by Slonczewski \cite{Slon} and
Berger. \cite{Berger1} Although the STT has many advantages over a magnetic
field in manipulating magnetization state, \cite{xrw0,xrw1} a large current
density is needed to achieve a technologically useful magnetization
switching speed, \cite{Yamaguchi,Parkin} but the associated Joule heating
could affect device performance. Therefore, the current density reduction
becomes a challenging issue from the spintronics application viewpoint. \cite%
{xrw2}

Many efforts have been devoted to the issue, including using optimized
time-dependent current pulse, \cite{xrw3} pure spin current, \cite{Lu} and
thermal activation. \cite{Hatami,Xia1} One direct approach is to increase
the magnitude of the STT under a given current through unique geometry
design. In 2003, Berger \cite{Berger2} proposed a novel magnetic multilayer
architecture called dual spin valve (DSV) where the free magnetic layer is
sandwiched between two thicker pinned magnetic layers with opposite
magnetizations. It is predicted that the STT applied on the free magnetic
layer should be much larger than that in a traditional spin valve for a
given current in the diffusion regime. \cite{Berger2} The argument is that
spins accumulate at both non-magnetic/magnetic interfaces of the free layer
and STT is proportional to spin accumulations. \cite{Berger2} However, it is
not clear whether this STT enhancement can occur in DSV in the ballistic
regime without spin accumulations. Moreover, the analytical formalism for
the STT in DSV is still an open problem up to now. \cite{You,Balaz} These
are the focuses of present paper. A full-quantum description of the STT,
valid when both the mean-free path and the spin-flip relaxation length are
larger than the thickness of the spacers, is presented. Averaged (over
electron phases) STT is obtained analytically within the Slonczewski's
semiclassical approach \cite{Slon} when all magnetic layers are perfect
polarizers. It is found that STT in DSV, depending on the orientation of
free layer and the degree of electron polarization, can be enhanced by a
factor of tens in comparison with that in a spin valve. The general cases of
arbitrary transmission and reflection coefficients of free layer are also
numerically calculated.

This paper is organized as follows: In Sec. II, a physical picture of STT
origin in both spin valve and DSV is presented. Sec. III is the theoretical
model and formulation of the electron transport through a DSV. The results,
including the analytical expression of the STT on the free layer within the
Slonczewski approach, are given in Sec. IV. Sec. V is the summary.

\section{PHYSICAL PICTURE}

\begin{figure}[tbph]
\begin{center}
\includegraphics[width=8.cm]{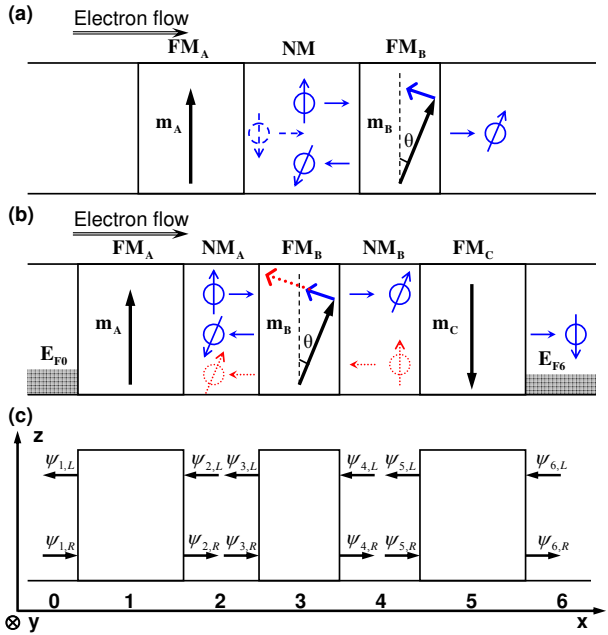}
\end{center}
\caption{(Color online) (a) Illustration of STT generated in spin valve. The
circles with arrows illustrate the electron flow with spin polarization
direction. (b) Schematic explanation of STT enhancement in DSV. $%
V_{b}=E_{F0}-E_{F6}$ is the applied bias. FM$_{A,B,C}$ are ferromagnetic
layers and NM$_{A,B}$ are nonmagnetic layers. (c) Wave functions at the
interfaces of regions ($0-6$) in DSV and the coordinates orientation.}
\end{figure}

It shall be useful to present first a physical picture about STT origin in
spin valve and, in particular, its enhancement in DSV. Consider a spin
valve, which is schematically shown in Fig. 1(a), consisting of a pinned
ferromagnetic layer FM$_{A}$ and a free ferromagnetic layer FM$_{B}$
separated by a nonmagnetic spacer NM. The magnetizations of FM$_A$ and FM$_B$
are represented by unit vectors, $\mathbf{m}_{A}$ and $\mathbf{m}_{B}$, and
their saturated magnetizations. For the simplicity, we assume that both
ferromagnetic layers are perfect spin filters, such that spins aligned
parallelly with the layer magnetization can completely transmit through the
layer, while antiparallel spins are totally reflected. In this ideal case, a
closed analytical solution of the STT can be obtained, which will be shown
later.

Consider electrons flowing from the left to the right in Fig. 1(a). The
right-going electrons are initially polarized along $\mathbf{m}_A$ direction
after passing through FM$_A$. They will remain their spin polarization when
they impinge on FM$_B$, as long as the spacer thickness is much shorter than
the spin-diffusion length, which is usually the case in nanoscale spin
valves. Because the polarization direction of FM$_B$ is noncollinear with
that of FM$_A$, an electron polarized along $\mathbf{m}_A$ is the
superposition of two states along $\mathbf{m}_B$ and $-\mathbf{m}_B$, so
that the component along $\mathbf{m}_B$ can transmit through FM$_B$ while
that along $-\mathbf{m}_B$ is totally reflected since FM$_B$ is a perfect
polarizer. Thus, there will be a net angular momentum transfer,
perpendicular to $\mathbf{m}_B$, from impinged electrons to FM$_B$,
resulting in a torque on FM$_B$ to align its magnetization toward $\mathbf{m}%
_A$, as shown by the blue arrow in Fig. 1(a). This is the origin of STT in
spin valve. It should be pointed out that the subsequent multiple
reflections of electrons within the NM spacer by the two FM/NM interfaces
will reduce the STT, since the reflected electrons (dashed symbols in Fig.
1(a)) from FM$_B$ to FM$_A$ and back to FM$_B$ will exert a torque along $-%
\mathbf{m}_A$. However, the net torque will not be zero since the reflected
electrons have smaller flux than that of the original injected electrons.

Let us now examine the spin transfer in a DSV shown in Fig. 1(b). On the top
of a usual spin valve (Fig. 1(a)), an additional pinned ferromagnetic layer
FM$_{C}$ with an antiparallelly aligned magnetization to FM$_{A}$, i.e., $%
\mathbf{m}_{C}=-\mathbf{m}_{A}$, is added so that the free layer FM$_{B}$ is
now sandwiched between FM$_{A}$ and FM$_{C}$, separated by two nonmagnetic
spacers NM$_{A}$ and NM$_{B}$, respectively. Similar to the case of a spin
valve, right-going electrons transmitting through FM$_{A}$ will exert a STT
on FM$_{B}$ to align $\mathbf{m}_{B}$ with $\mathbf{m}_{A}$, as shown by the
blue arrow in Fig. 1(b). After the electrons transmit through FM$_{B}$, most
of them will be reflected by FM$_{C}$ and then impinge again on FM$_{B}$, as
shown by red-dashed symbols in Fig. 1(b). Thus, they will exert another STT
on FM$_{B}$ along $-\mathbf{m}_{C}=\mathbf{m}_{A}$ direction, resulting in
the STT enhancement. Multiple reflections in region 2 (Fig. 1(c)) tend to
reduce the torque but will not cancel it totally. In the following sections
we will verify this physical picture through a full-quantum mechanics
calculation.

\section{MODEL AND FORMULATION}

Single charge and spin transport theory in magnetic multilayers was well
developed and approaches vary from classical Valet-Fert theory in the
diffusive regime, \cite{Valet} matrix Boltzmann equation formalism, \cite%
{Stiles2} to full-quantum mechanical treatments. \cite%
{Slon,Stiles1,Waldron,Brataas1,Krstajic,Waintal,Xiao} In the present paper,
we will adopt a full-quantum mechanical description called scattering matrix
method \cite{Krstajic,Waintal,Xiao} that is valid for ballistic
transmission. We assume that interfaces are flat and clean and all spin-flip
processes are negligible so that momentum $\mathbf{k}$ is a good quantum
number in each layer. Wavefunction at the interface of layers labeled by $%
n=1-6$ as shown in Fig. 1(c) can be written as a two-component spinor
multiplied by spatial plane wave,
\begin{equation}
\psi _{n,\alpha }\left( x\right) =\binom{\chi _{n,\alpha ,\uparrow }}{\chi
_{n,\alpha ,\downarrow }}e^{ix\left( k_{x}\right) _{n,\alpha }},
\end{equation}%
where $\alpha $ denotes propagation directions, $\alpha =L$ for leftward and
$\alpha =R$ for rightward. $k_{x}$ is the $x-$component of wave vector. $%
\chi _{n,\alpha ,\uparrow (\downarrow )}$ is the spin-up (-down) probability
amplitude. $z-$axis is along $\mathbf{m}_{A}$. $\mathbf{m}_{B}$ is specified
by a polar angle $\theta $ and an azimuthal angle $\phi $.

Incoming and outgoing spinor states in each region are connected to each
other by a scattering matrix. \cite{Krstajic,Waintal,Xiao} In region $n$ $%
(=1,2,3,4,5),$ $\psi _{n,L}$ and $\psi _{n+1,R}$ are outgoing spinors while $%
\psi _{n,R}$ and $\psi _{n+1,L}$ are incoming spinors. They relate to each
other by a scattering matrix $\hat{S}_{n}$,
\begin{equation}
\left(
\begin{array}{c}
\psi _{n,L} \\
\psi _{n+1,R}%
\end{array}%
\right) =\hat{S}_{n}\left(
\begin{array}{c}
\psi _{n,R} \\
\psi _{n+1,L}%
\end{array}%
\right) .
\end{equation}%
Note that $\psi _{6,L}=0$ since electrons flow from the left reservoir to
the right one. $\hat{S}_{n}$ is a $4\times 4$ matrix and can be expressed by
transmission and reflection coefficients of each scattering region,
\begin{equation}
\hat{S}_{n}=\left(
\begin{array}{cc}
\hat{r}_{n} & \hat{t}_{n} \\
\hat{t}_{n} & \hat{r}_{n}%
\end{array}%
\right) ,
\end{equation}%
where $\hat{t}_{n}\left( \hat{r}_{n}\right) $ with $n=1,2,3,4,5$ are $%
2\times 2$ transmission (reflection) matrices for layers FM$_{A}$, NM$_{A}$,
FM$_{B}$, NM$_{B}$, and FM$_{C}$, respectively.

We will first treat the pinned ferromagnetic layers FM$_A$ and FM$_C$ as
perfect polarizers. Hence $\hat{t}_1,$ $\hat{r}_1,$ $\hat{t}_5,$ and $\hat{r}%
_5$ take the following forms
\begin{equation}
\hat{t}_1 = \hat{r}_5= \left(
\begin{array}{cc}
1 & 0 \\
0 & 0%
\end{array}%
\right) ,\text{ }\hat{r}_1=\hat{t}_5= \left(
\begin{array}{cc}
0 & 0 \\
0 & 1%
\end{array}%
\right) .
\end{equation}
Since there is no scatterings in nonmagnetic layers NM$_A$ and NM$_B$ and
the propagation of electrons in these layers accumulate dynamical phases, we
have
\begin{equation}
\hat{t}_{2}=\exp \left( i\varphi _{A}\right) \hat{I},\text{ }\hat{t}%
_{4}=\exp \left( i\varphi _{B}\right) \hat{I},\text{ }\hat{r}_{2}=\hat{r}%
_{4}=0,
\end{equation}%
where $\varphi _{A}$ and $\varphi _{B}$ are the corresponding phase shifts
and $\hat{I}$ is the $2\times 2$ unit matrix. The scattering matrix for FM$%
_{B}$ can be expressed by the angles $\theta $ and $\phi $ as
\begin{eqnarray}
\hat{t}_{3} &=&\hat{R}\left( \theta ,\phi \right) \mathbf{t}\hat{R}%
^{-1}\left( \theta ,\phi \right) ,  \notag \\
\hat{r}_{3} &=&\hat{R}\left( \theta ,\phi \right) \mathbf{r}\hat{R}%
^{-1}\left( \theta ,\phi \right) ,
\end{eqnarray}%
where $\hat{R}\left( \theta ,\phi \right)$ is the rotation that brings $\hat
z$ into $\mathbf{m}_{B}$,
\begin{equation}
\hat{R}=e^{-\frac{i\phi }{2}\sigma _{z}}e^{-\frac{i\theta }{2}\sigma
_{y}}=\left(
\begin{array}{cc}
e^{-\frac{i\phi }{2}}\cos \frac{\theta }{2} & -e^{-\frac{i\phi }{2}}\sin
\frac{\theta }{2} \\
e^{\frac{i\phi }{2}}\sin \frac{\theta }{2} & e^{\frac{i\phi }{2}}\cos \frac{%
\theta }{2}%
\end{array}%
\right) .
\end{equation}
$\mathbf{t}$ and $\mathbf{r}$ are the transmission and reflection matrices
when $\mathbf{m}_B$ is chosen as the quantization axis
\begin{equation}
\mathbf{t}=\left(
\begin{array}{cc}
t_{u} & 0 \\
0 & t_{d}%
\end{array}%
\right) ,\text{ }\mathbf{r}=\left(
\begin{array}{cc}
r_{u} & 0 \\
0 & r_{d}%
\end{array}%
\right) ,
\end{equation}%
where $t_{u},$ $t_{d},$ $r_{u},$ and $r_{d}$ are transmission and reflection
parameters. Subscripts $u$ and $d$ stand for spin-up (majority) and
spin-down (minority), respectively. These parameters are complex numbers in
general. $t_u=1,$ $t_d=0$ and $r_u=0,$ $r_d=1$ if FM$_B$ is a perfect
polarizer.

To find the STT on free layer FM$_{B}$, we need to obtain the scattering
states at the two interfaces of FM$_{B}$. The spin-dependent scattering wave
functions can be expressed in terms of incoming wave $\psi _{1,R},$ such as $%
\psi _{n,\alpha }=\hat{P}_{n,\alpha }\psi _{1,R}$ $\left( n=3,4,6\right) ,$
where matrices $\hat{P}_{n,\alpha }$ are given by
\begin{eqnarray}
\hat{P}_{3,R} &=&\Bigl [\left( \hat{I}-\hat{t}_{2}^{2}\hat{r}_{1}\hat{r}%
_{3}\right) -\hat{t}_{2}^{2}\hat{t}_{4}^{2}\hat{r}_{1}\hat{t}_{3}\hat{r}_{5}%
\Bigl (\hat{I}-\hat{t}_{4}^{2}\hat{r}_{3}\hat{r}_{5}\Bigr )^{-1}\hat{t}_{3}%
\Bigr ]^{-1}  \notag \\
&&\hat{t}_{2}\hat{t}_{1},  \notag \\
\hat{P}_{3,L} &=&\hat{r}_{3}\hat{P}_{3,R}+\hat{t}_{4}\hat{t}_{3}\hat{r}_{5}%
\hat{Q},  \notag \\
\hat{P}_{4,R} &=&\hat{t}_{4}^{-1}\hat{Q},\text{ }\hat{P}_{4,L}=\hat{t}_{4}%
\hat{r}_{5}\hat{Q},  \notag \\
\hat{P}_{6,R} &=&\hat{t}_{5}\hat{Q},
\end{eqnarray}%
with
\begin{eqnarray}
\hat{Q} &=&\Bigl [\left( \hat{I}-\hat{t}_{4}^{2}\hat{r}_{3}\hat{r}%
_{5}\right) -\hat{t}_{2}^{2}\hat{t}_{4}^{2}\hat{t}_{3}\Bigl (\hat{I}-\hat{t}%
_{2}^{2}\hat{r}_{1}\hat{r}_{3}\Bigr )^{-1}\hat{r}_{1}\hat{t}_{3}\hat{r}_{5}%
\Bigr ]^{-1}  \notag \\
&&\hat{t}_{3}\Bigl (\hat{I}-\hat{t}_{2}^{2}\hat{r}_{1}\hat{r}_{3}\Bigr )^{-1}%
\hat{t}_{2}\hat{t}_{4}\hat{t}_{1}.
\end{eqnarray}

After some algebras, we have
\begin{subequations}
\begin{eqnarray}
&&\hat{Q}=\frac{\sqrt{z_{1}z_{2}}}{D\left( z_{1},z_{2}\right) }\left[
\begin{array}{cc}
Q^{\uparrow \uparrow }\left( z_{1},z_{2}\right) & 0 \\
Q^{\downarrow \uparrow }\left( z_{1},z_{2}\right) & 0%
\end{array}%
\right] ,  \label{Q} \\
&&\hat{P}_{3,R}=\frac{\sqrt{z_{1}}}{D\left( z_{1},z_{2}\right) }\left[
\begin{array}{cc}
P_{3,R}^{\uparrow \uparrow }\left( z_{1},z_{2}\right) & 0 \\
P_{3,R}^{\downarrow \uparrow }\left( z_{1},z_{2}\right) & 0%
\end{array}%
\right] ,  \label{P1} \\
&&\hat{P}_{3,L}=\frac{\sqrt{z_{1}}}{D\left( z_{1},z_{2}\right) }\left[
\begin{array}{cc}
P_{3,L}^{\uparrow \uparrow }\left( z_{1},z_{2}\right) & 0 \\
P_{3,L}^{\downarrow \uparrow }\left( z_{1},z_{2}\right) & 0%
\end{array}%
\right] ,  \label{P2} \\
&&\hat{P}_{4,R}=\frac{\sqrt{z_{1}}}{D\left( z_{1},z_{2}\right) }\left[
\begin{array}{cc}
Q^{\uparrow \uparrow }\left( z_{1},z_{2}\right) & 0 \\
Q^{\downarrow \uparrow }\left( z_{1},z_{2}\right) & 0%
\end{array}%
\right] , \\
&&\hat{P}_{4,L}=\frac{z_{2}\sqrt{z_{1}}}{D\left( z_{1},z_{2}\right) }\left[
\begin{array}{cc}
Q^{\uparrow \uparrow }\left( z_{1},z_{2}\right) & 0 \\
0 & 0%
\end{array}%
\right] ,
\end{eqnarray}%
where $z_{1}=\exp \left( i2\varphi _{A}\right) ,\ z_{2}=\exp \left(
i2\varphi _{B}\right) ,$ and
\end{subequations}
\begin{subequations}
\begin{eqnarray}
&&Q^{\uparrow \uparrow }\left( z_{1},z_{2}\right) =t_{u}\cos ^{2}\frac{%
\theta }{2}+t_{d}\sin ^{2}\frac{\theta }{2}  \notag \\
&&-z_{1}\left( r_{d}t_{u}\cos ^{2}\frac{\theta }{2}+r_{u}t_{d}\sin ^{2}\frac{%
\theta }{2}\right) , \\
&&Q^{\downarrow \uparrow }\left( z_{1},z_{2}\right) =\frac{1}{2}e^{i\phi
}\sin \theta \Bigl [t_{u}-t_{d}  \notag \\
&&+\left( z_{1}+z_{2}\right) \left( r_{u}t_{d}-r_{d}t_{u}\right) \\
&&+z_{1}z_{2}\Bigl (%
t_{u}r_{d}^{2}+t_{d}t_{u}^{2}-t_{u}t_{d}^{2}-t_{d}r_{u}^{2}\Bigr )\Bigr ],
\notag \\
&&D\left( z_{1},z_{2}\right) =1-z_{1}\left( r_{u}\sin ^{2}\frac{\theta }{2}%
+r_{d}\cos ^{2}\frac{\theta }{2}\right)  \notag \\
&&-z_{2}\left( r_{u}\cos ^{2}\frac{\theta }{2}+r_{d}\sin ^{2}\frac{\theta }{2%
}\right) + \\
&&z_{1}z_{2}\Bigl [r_{u}r_{d}+\frac{\sin ^{2}\theta }{4}\Bigl (\left(
r_{u}-r_{d}\right) ^{2}-\left( t_{u}-t_{d}\right) ^{2}\Bigr )\Bigr ],  \notag
\\
&&P_{3,R}^{\uparrow \uparrow }\left( z_{1},z_{2}\right) =D\left(
z_{1},z_{2}\right) , \\
&&P_{3,R}^{\downarrow \uparrow }\left( z_{1},z_{2}\right) =\frac{1}{2}%
z_{1}e^{i\phi }\sin \theta \Bigl [r_{u}-r_{d}  \notag \\
&&+z_{2}\left( -r_{u}^{2}+r_{u}r_{d}+t_{u}^{2}-t_{u}t_{d}\right) \cos ^{2}%
\frac{\theta }{2} \\
&&+z_{2}\left( r_{d}^{2}-r_{u}r_{d}-t_{d}^{2}+t_{u}t_{d}\right) \sin ^{2}%
\frac{\theta }{2}\Bigr ],  \notag \\
&&P_{3,L}^{\uparrow \uparrow }\left( z_{1},z_{2}\right) =z_{2}\left(
t_{u}\cos ^{2}\frac{\theta }{2}+t_{d}\sin ^{2}\frac{\theta }{2}\right) ^{2}
\notag \\
&&-z_{2}\left( r_{u}\cos ^{2}\frac{\theta }{2}+r_{d}\sin ^{2}\frac{\theta }{2%
}\right) ^{2}  \notag \\
&&+r_{u}\cos ^{2}\frac{\theta }{2}+r_{d}\sin ^{2}\frac{\theta }{2} \\
&&+z_{1}z_{2}\Bigl [r_{u}\left( r_{d}^{2}-t_{d}^{2}\right) \sin ^{2}\frac{%
\theta }{2}+r_{d}\left( r_{u}^{2}-t_{u}^{2}\right) \cos ^{2}\frac{\theta }{2}%
\Bigr ],  \notag \\
&&P_{3,L}^{\downarrow \uparrow }\left( z_{1},z_{2}\right) =\frac{%
P_{3,R}^{\downarrow \uparrow }\left( z_{1},z_{2}\right) }{z_{1}}.
\end{eqnarray}%
The notation $X^{\downarrow \uparrow }$ $(X=\hat{Q},\hat{P}_{3,R},\hat{P}%
_{3,L})$ refers to the transition amplitude from spin-up to spin-down states.

\section{RESULTS AND DISCUSSIONS}

\subsection{Charge current}

An applied bias $V_{b}$ shown in Fig. 1(b) generates a charge current $J_{e}$
and a spatially dependent spin current $\mathbf{J}_{s}$ through the device.
At zero temperature, the charge current reads
\end{subequations}
\begin{equation}
J_{e}=\int dE\sum\limits_{\mathbf{q}}j_{e}\left( \mathbf{q}\right) ,
\end{equation}%
with charge current density%
\begin{equation}
j_{e}\left( \mathbf{q}\right) =e\frac{\hslash k_{x}}{m}\psi _{6,R}^{\dag
}\psi _{6,R},
\end{equation}%
where $\mathbf{q}$ is the transverse wave vector with energy $E$, $k_{x}^{2}+%
\mathbf{q}^{2}=2mE/\hslash ^{2},$ and $e$, $m$, $\hslash $ are the electron
charge, electron mass and the Planck constant. The charge current density
can also be written as
\begin{equation}
j_{e}=e\frac{\hslash k_{x}}{m}T_{e}\left( z_{1},z_{2}\right) ,
\label{current}
\end{equation}%
with transmission coefficient
\begin{equation}
T_{e}\left( z_{1},z_{2}\right) =\frac{\left\vert Q^{\downarrow \uparrow
}\left( z_{1},z_{2}\right) \right\vert ^{2}}{\left\vert D\left(
z_{1},z_{2}\right) \right\vert ^{2}}.  \label{Tc}
\end{equation}

In the case that electrons propagate ballistically through NM$_A$ and NM$_B$%
, the phase shifts in normal metals are given by $\varphi _A=k_xl_A$ and $%
\varphi _B=k_xl_B$ where $l_A$ and $l_B$ are the widths of NM$_A$ and NM$_B,$
respectively. For sufficiently thick (much bigger than electron Fermi wave
length but still smaller than the spin diffusion length) NM layers, $%
\varphi_A$ and $\varphi_B$ vary rapidly from state to state. Thus, when one
sums up contributions from different electronic states (different $k_x$), it
is justifiable to assume $\varphi_A$ and $\varphi_B$ to be random, \cite%
{Waintal} and $z_1=\exp\left( i2\varphi_A\right)$ and $z_2=\exp\left(i2%
\varphi_B\right)$ are equally distributed on the unit circle of the complex
plane. \cite{Krstajic, Waintal} However, one should note that $z_1$ and $z_2$
are not independent under the ballistic assumption since $z_2=\left(
z_1\right)^p$ with $p=l_B/l_A.$ The average transmission coefficient is then
\begin{eqnarray}
\left\langle T_e\right\rangle &=&\frac{1}{\pi}\int_0^\pi T_e
\left(\varphi_A,p\varphi_A\right) d\varphi_A  \notag \\
&=&\frac{1}{2\pi i}\oint_C\frac{T_e\left( z_1,\left( z_1\right) ^p\right) }{%
z_1}dz_1,  \label{complex}
\end{eqnarray}%
where $C$ is contour $\left\vert z_1\right\vert =1.$ The contour integral
for $p=1$, corresponding to the symmetric DSV configuration, \cite%
{Balaz,Gmitra} is
\begin{equation}
\left\langle T_{e}\right\rangle =\sum_{l=1}^{s}\text{Res}\left( \frac{%
T_{e}\left( z_{1}\right) }{z_{1}};z_{1,l}\right) ,  \label{Res0}
\end{equation}%
where $z_{1,l}$ is the $l-$th pole of function $T_{e}\left( z_{1}\right)
/z_{1}$ and $s$ is the total number of poles inside the unit circle. In case
when FM$_B$ is perfect, i.e., $t_u=1$ and $t_d=0$ ($r_u=0,$ and $r_d=1$),
function $T_e\left( z_1\right) /z_1$ has only a second-order pole $z_1=0$
inside the unit circle. Thus, we can get the average transmission%
\begin{equation}
\left\langle T_{e}\right\rangle \left( t_{u}=1,\text{ }t_{d}=0\right) =\frac{%
\sin ^{2}\theta }{2}.
\end{equation}
\begin{figure}[tbph]
\begin{center}
\includegraphics[width=8.0cm]{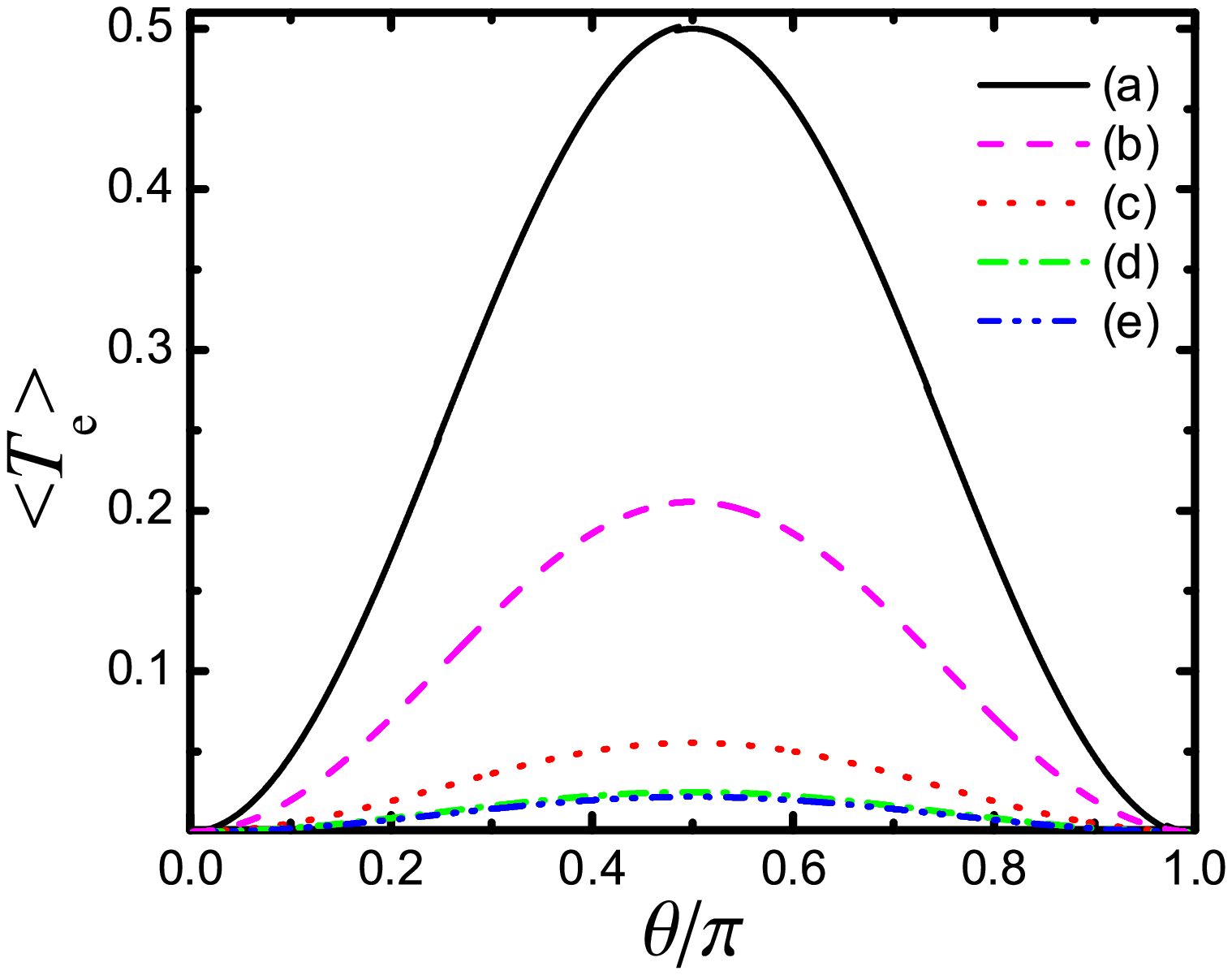}
\end{center}
\caption{(Color online) Average transmission coefficient versus angle $%
\protect\theta $ for (a) $t_u=1$ and $t_d=0$; (b) $\left\vert
t_u\right\vert^2=0.84$ and $\left\vert t_d\right\vert ^2=0.17$; (c) $%
\left\vert t_u\right\vert ^2=0.79$ and $\left\vert t_d\right\vert ^2=0.49$;
(d) $\left\vert t_u\right\vert ^2=0.66$ and $\left\vert t_d\right\vert
^2=0.44$; and (e) $\left\vert t_u\right\vert ^2=0.73$ and $\left\vert
t_d\right\vert ^2=0.54$.}
\end{figure}

Figure 2 is the average transmission coefficient versus angle $\theta $ for
(a) perfect polarizers; (b) (001) interface of Au/Fe; (c) (001) interface of
Cu/Co; (d) (110) interface of Cu/Co; (e) (111) interface of Cu/Co. The model
parameters for (b-e) are obtained from Ref. \onlinecite{Stiles3}, where they
were extracted from the first-principles calculations. All transmission and
reflection amplitudes are assumed to be real. We find that the average total
transmission approach zero as $\theta $ goes to $0$ or $\pi $ even if $t_{d}$
is finite for the minority electrons, which is different from the result
(Fig. 3 in Ref. \onlinecite{Krstajic}) in traditional spin valve. The
reasons are as follows. For $\theta =0$, all electrons are polarized along $%
\mathbf{m}_{A}$ after passing through FM$_{B}$. They will be totally
reflected by FM$_{C}$ because of $\mathbf{m}_{C}=-\mathbf{m}_{A}$, leading
to zero electric current. On the other hand, for $\theta =\pi $, all
electrons that transmit FM$_{A}$ will be completely reflected by FM$_{B}$ if
it is a perfect polarizer because electron spins are antiparallel to $%
\mathbf{m}_{B}$, while electrons passing through FM$_{B}$ will maintain
polarization along $\mathbf{m}_{A}$ in the case of $t_{d}\neq 0$ and they
will be totally reflected by FM$_{C}$ because of $\mathbf{m}_{C}=-\mathbf{m}%
_{A}$. Hence, the average transmission in DSV vanishes at $\theta =0$ or $%
\pi .$

The total charge current flowing through the DSV can be obtained by summing
Eq. \eqref{current}\ over the transverse momentum $\mathbf{q}$. To find an
analytical expression, we will adopt the semiclassical Slonczewski approach.
\cite{Slon} Within the Stoner description of magnetism and let $\Delta$ be
the exchange energy of two spin bands of FM$_B$, one can define two Fermi
wave vectors $K_{+}$ and $K_-$ for majorities and minorities,
\begin{equation}
K_+=\sqrt{2mE/\hslash^2},\text{ }K_-=\sqrt{2m\left( E-\Delta\right) /\hslash
^2}.
\end{equation}%
For a ferromagnetic metal, we assume that the Fermi energy lies above the
exchange potential, and electrons in NMs are ideally matched with the
majority electrons in FM, i.e., $k=K_{+}.$ The possible momentum states that
contribute to the current can be divided into three ranges. \cite%
{Slon,Krstajic}

\emph{Range a}: $0\leq q<K_{-}.$ Electrons of both spins in these states
contribute to charge current $J_a$,
\begin{eqnarray}
J_{a} &=&2e\frac{\hslash }{\left( 2\pi \right) ^{2}m}\int_{0}^{K_{-}}\sqrt{%
K_{+}^{2}-q^{2}}qdq  \notag \\
&=&\frac{2}{3}e\frac{\hslash }{\left( 2\pi \right) ^{2}m}\left[ \left(
K_{+}^{2}\right) ^{3/2}-\left( K_{+}^{2}-K_{-}^{2}\right) ^{3/2}\right] .
\end{eqnarray}

\emph{Range b}: $K_{-}\leq q<K_{+}.$ Only majority spin electrons contribute
to current $J_b$,
\begin{eqnarray}
J_{b} &=&e\frac{\hslash }{\left( 2\pi \right) ^{2}m}\int_{K_{-}}^{K_{+}}%
\sqrt{K_{+}^{2}-q^{2}}qdq  \notag \\
&=&\frac{1}{3}e\frac{\hslash }{\left( 2\pi \right) ^{2}m}\left(
K_{+}^{2}-K_{-}^{2}\right) ^{3/2}.
\end{eqnarray}

\emph{Range c}: $K_{+}\leq q.$ All electrons are totally reflected, and
there is no charge current flow, i.e., $J_{c}=0.$

Using the conventional definition of spin polarization
\begin{equation}
P=\frac{n_{+}-n_{-}}{n_{+}+n_{-}}=\frac{K_{+}-K_{-}}{K_{+}+K_{-}},
\end{equation}%
where $n_{\pm }$ are the majority/minority spin densities at Fermi level in
the FMs, the ratio $J_{a}/J_{b}$ can be written as a function of
polarization $P$,
\begin{equation}
\frac{J_{a}}{J_{b}}=\frac{\left( 1+P\right) ^{3}}{4P^{3/2}}-2.
\end{equation}

Note that $J_{b}$\ is the maximal polarized current for parallel FMs
configuration. Then, to get the total charge current, it should be
multiplied by the average transmission coefficient. Thus, the total electron
current is given by \cite{Slon,Krstajic}%
\begin{equation}
J_{e}=J_{a}+\left\langle T_{e}\right\rangle J_{b}.
\end{equation}

\subsection{Spin current and spin transfer torque}

The spin currents on two sides of FM$_{B}$ are
\begin{eqnarray}
\mathbf{J}_{3s} &=&\int dE\sum\limits_{\mathbf{q}}\mathbf{j}_{3s}\left(
\mathbf{q}\right) ,  \label{s3} \\
\mathbf{J}_{4s} &=&\int dE\sum\limits_{\mathbf{q}}\mathbf{j}_{4s}\left(
\mathbf{q}\right) ,  \label{s4}
\end{eqnarray}%
with spin current densities%
\begin{eqnarray}
\mathbf{j}_{3s}\left( \mathbf{q}\right) &\mathbf{=}&\frac{\hslash ^{2}k_{x}}{%
2m}\left( \psi _{3,R}^{\dag }\mathbf{\hat{\sigma}}\psi _{3,R}-\psi
_{3,L}^{\dag }\mathbf{\hat{\sigma}}\psi _{3,L}\right) , \\
\mathbf{j}_{4s}\left( \mathbf{q}\right) &\mathbf{=}&\frac{\hslash ^{2}k_{x}}{%
2m}\left( \psi _{4,R}^{\dag }\mathbf{\hat{\sigma}}\psi _{4,R}-\psi
_{4,L}^{\dag }\mathbf{\hat{\sigma}}\psi _{4,L}\right) ,
\end{eqnarray}%
where $\mathbf{\hat{\sigma}}=\left( \sigma _{x},\sigma _{y},\sigma
_{z}\right) $ are Pauli matrices. It is convenient to recast $\mathbf{\hat{%
\sigma}}$ in local orthogonal coordinates of $\mathbf{m}_{B}\times \left(
\mathbf{m}_{B}\times \mathbf{m}_{A}\right) \mathbf{,}$ $\mathbf{m}_{B}\times
\mathbf{m}_{A},$ and $\mathbf{m}_{B}$ such that
\begin{equation}
\mathbf{\hat{\sigma}}\mathbf{=}\sigma _{1}\mathbf{m}_{B}\times \left(
\mathbf{m}_{B}\times \mathbf{m}_{A}\right) +\sigma _{2}\mathbf{m}_{B}\times
\mathbf{m}_{A}+\sigma _{3}\mathbf{m}_{B},  \label{local coordinates}
\end{equation}%
with
\begin{eqnarray}
\sigma _{1} &=&\frac{1}{\sin \theta }\left(
\begin{array}{cc}
-\sin \theta & \cos \theta e^{-i\phi } \\
\cos \theta e^{i\phi } & \sin \theta%
\end{array}%
\right) ,  \notag \\
\sigma _{2} &=&\frac{1}{\sin \theta }\left(
\begin{array}{cc}
0 & ie^{-i\phi } \\
-ie^{i\phi } & 0%
\end{array}%
\right) ,  \notag \\
\sigma _{3} &=&\left(
\begin{array}{cc}
\cos \theta & \sin \theta e^{-i\phi } \\
\sin \theta e^{i\phi } & -\cos \theta%
\end{array}%
\right) .
\end{eqnarray}

The STT on FM$_{B}$ is equal to the difference of the spin currents on both
sides of the ferromagnet,
\begin{equation}
\boldsymbol{\Gamma }=\mathbf{J}_{3s}-\mathbf{J}_{4s},
\end{equation}%
and the STT density is%
\begin{equation}
\boldsymbol{\tau }=\mathbf{j}_{3s}-\mathbf{j}_{4s}.
\end{equation}%
Thus, we have
\begin{equation}
\boldsymbol{\tau }=a_{1}\mathbf{m}_{B}\times \left( \mathbf{m}_{B}\times
\mathbf{m}_{A}\right) +a_{2}\mathbf{m}_{B}\times \mathbf{m}_{A}+a_{3}\mathbf{%
m}_{B},
\end{equation}%
where
\begin{eqnarray}
a_{i} &=&\frac{\hslash ^{2}k_{x}}{2m}\Bigl (\psi _{3,R}^{\dag }\sigma
_{i}\psi _{3,R}+\psi _{4,L}^{\dag }\sigma _{i}\psi _{4,L}  \notag \\
&&-\psi _{3,L}^{\dag }\sigma _{i}\psi _{3,L}\mathbf{-}\psi _{4,R}^{\dag
}\sigma _{i}\psi _{4,R}\Bigr ),\text{ }\left( i=1,2,3\right) .
\label{a-term}
\end{eqnarray}

First of all, we can show $a_{3}=0$ because of the particle current
conservation and the absence of spin-flipping. This can be understood as
follows: We first rotate $\hat{z}$ to $\mathbf{m}_{B},$ then spinor state $%
\psi _{n,\alpha }=\hat{R}\left( \theta ,\phi \right) \tilde{\psi}_{n,\alpha
} $ where $\tilde{\psi}_{n,\alpha }$\ is the electronic state $seen$ along $%
\mathbf{m}_{B}.$ Then, each spin density term $\psi _{n,\alpha }^{\dag
}\sigma _{3}\psi _{n,\alpha }$ in Eq. \eqref{a-term} becomes $\tilde{\psi}%
_{n,\alpha }^{\dag }\hat{R}^{\dagger }\left( \theta ,\phi \right) \sigma _{3}%
\hat{R}\left( \theta ,\phi \right) \tilde{\psi}_{n,\alpha }=\tilde{\psi}%
_{n,\alpha }^{\dag }\sigma _{z}\tilde{\psi}_{n,\alpha },$ so that spin up
state (parallel to $\mathbf{m}_{B},$) and spin down state (anti-parallel to $%
\mathbf{m}_{B}$) are decoupled without mixing. In the absence of
spin-flipping, both spin-up and spin-down particle currents are conserved.
Thus, the STT projected along local magnetization $\mathbf{m}_{B}$ is zero.
Here we have used identity $\hat{R}^{\dagger }\left( \theta ,\phi \right)
\sigma _{3}\hat{R}\left( \theta ,\phi \right) =\sigma _{z}.$ The physical
consequence is that STT can only rotate the magnetization without change its
magnitude.

Therefore, the STT can be divided into an in-plane (Slonczewski) term \cite%
{Slon} $\boldsymbol{\tau}_\parallel=a_1\mathbf{m}_B\times \left( \mathbf{m}%
_B\times \mathbf{m}_A\right)$ and an out-of-plane (field-like) term \cite%
{Zhang} $\boldsymbol{\tau}_\perp=a_2\mathbf{m}_B\times \mathbf{m}_A.$ We
note that the out-of-plane torque will vanish if $t_{u},$ $t_{d},$ $r_{u},$
and $r_{d}$ are real. This is because all the spins of both transmitted and
reflected electrons are in the same plane spanned by $\mathbf{m}_A$ and $%
\mathbf{m}_B$ under the condition, and they can only vary in this plane.

Parameters $a_{1}$ and $a_{2}$ are important to understand the spin dynamics
in DSV. \cite{You,Balaz} In Ref. \onlinecite{You}, these parameters are
chosen to vary continuously without geometry dependence, while they are
calculated in the diffusice transport limit in Ref. \onlinecite{Balaz}. To
find $a_{1}$ and $a_{2}$ in our model, one needs to compute quantities
\begin{eqnarray}
T_{\sigma _{i}}\left( z_{1}\right)  &=&\frac{1}{2}\text{Tr}\Bigl (\hat{P}%
_{3,R}^{\dag }\sigma _{i}\hat{P}_{3,R}-\hat{P}_{3,L}^{\dag }\sigma _{i}\hat{P%
}_{3,L}-\hat{P}_{4,R}^{\dag }\sigma _{i}\hat{P}_{4,R}  \notag \\
&&+\hat{P}_{4,L}^{\dag }\sigma _{i}\hat{P}_{4,L}\Bigr ),\text{ }\left(
i=1,2\right) ,
\end{eqnarray}%
since $a_{i}=T_{\sigma _{i}}\left( z_{1}\right) (\hslash ^{2}k_{x})/m,$ $%
\left( i=1,2\right) $. Perform the same averaging procedure as we did on the
charge current early, one finds
\begin{equation}
\left\langle T_{\sigma _{i}}\right\rangle =\sum_{l_{i}=1}^{s_{i}}\text{Res}%
\left( \frac{T_{\sigma _{i}}\left( z_{1}\right) }{z_{1}};z_{1,l_{i}}\right) ,%
\text{ }\left( i=1,2\right) ,  \label{Res1}
\end{equation}%
where $z_{1,l_{i}}$ is the $l_{i}-$th pole of function $T_{\sigma
_{i}}\left( z_{1}\right) /z_{1}$ inside the unit circle in the complex plane
with $s_{i}$ the corresponding total pole number.

Since only region \emph{b} contributes to the spin current, \cite%
{Slon,Krstajic} one has $\left\langle a_{i}\right\rangle =\frac{\hslash }{e}%
\left\langle T_{\sigma _{i}}\right\rangle J_{b},$ $\left( i=1,2\right) .$
Thus the average STT on the free magnetic layer is
\begin{equation}
\mathbf{\Gamma =}g_{1}\left( \theta \right) \frac{\hslash }{e}J_{e}\mathbf{m}%
_{B}\times \left( \mathbf{m}_{B}\times \mathbf{m}_{A}\right) +g_{2}\left(
\theta \right) \frac{\hslash }{e}J_{e}\mathbf{m}_{B}\times \mathbf{m}_{A},
\label{STT}
\end{equation}%
with scalar functions
\begin{equation}
g_{i}\left( \theta \right) =\frac{\left\langle T_{\sigma _{i}}\right\rangle
}{J_{a}/J_{b}+\left\langle T_{e}\right\rangle },\text{ }\left( i=1,2\right) .
\label{g1}
\end{equation}

The STT $\mathbf{\Gamma }$ consists of two terms. The first one is
Slonczewski torque $\mathbf{\Gamma }_{\parallel }\mathbf{=}g_{1}\left(
\theta \right) \frac{\hslash }{e}J_{e}\mathbf{m}_{B}\times \left( \mathbf{m}%
_{B}\times \mathbf{m}_{A}\right) .$ And the second one is field-like torque $%
\mathbf{\Gamma }_{\perp }\mathbf{=}g_{2}\left( \theta \right) \frac{\hslash
}{e}J_{e}\mathbf{m}_{B}\times \mathbf{m}_{A}.$

\begin{figure}[tbph]
\begin{center}
\includegraphics[width=8.0cm]{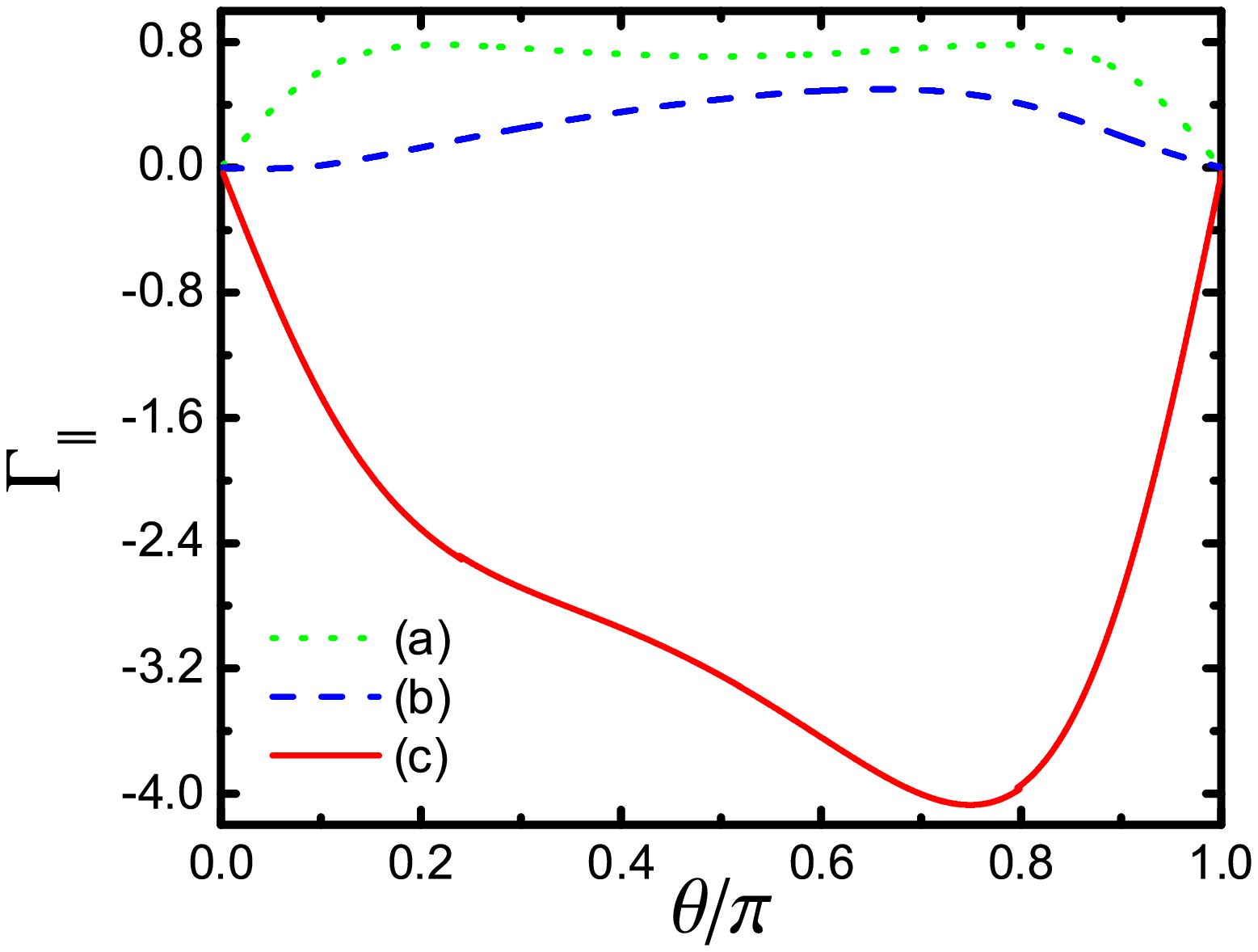}
\end{center}
\caption{(Color online) Slonczewski torque ${\Gamma }_{\parallel }$ versus $%
\protect\theta $ for polarization $P=0.6$ under three different conditions:
(a) $t_{u}=1$ and $t_{d}=0 $; (b) $\left\vert t_{u}\right\vert ^{2}=0.99$
and $\left\vert t_{d}\right\vert ^{2}=0.1$; and (c) $\left\vert
t_{u}\right\vert ^{2}=0.84$ and $\left\vert t_{d}\right\vert ^{2}=0.17$. The
unit of torque is $\frac{\hslash }{e}J_{e}$.}
\end{figure}

The general analytical forms of $g_{1}\left( \theta \right) $ and $%
g_{2}\left( \theta \right) $ are difficult to find because of the
complicated residue calculations in Eqs. \eqref{Res0} and \eqref{Res1}.
However, they can be numerically evaluated for any given material with
definite material parameters $t_{u},$ $t_{d},$ $r_{u},$ and $r_{d}.$ In Fig.
3, we present the numerical calculations of the magnitude of Slonczewski
torque $\Gamma _{\parallel }\mathbf{=-}g_{1}\left( \theta \right) \frac{%
\hslash }{e}J_{e}\sin \theta $ per unit current versus angle $\theta $ at
polarization coefficient $P=0.6$ under three different conditions, which
shows that different transmission probabilities have strong impact on the
torque and may change the sign of the torque. Similar sign reversal of STT
is demonstrated in usual spin valve. \cite{Waintal}

Here, in order to directly compare our results in DSV with Slonczewski's
result in conventional spin valve, \cite{Slon} let us consider the case of
ideal FM$_{B}$, i.e., $t_{u}=1$ and $t_{d}=0$ $\left( r_{u}=0,r_{d}=1\right)
$. After some algebras, we obtain $T_{\sigma _{1}}=\frac{1}{4}\left(
z_{1}+z_{1}^{\ast }-2\right) $ and $T_{\sigma _{2}}\left( z_{1}\right) =-%
\frac{i}{4}\left( z_{1}-z_{1}^{\ast }\right) $. The averaged values are then
\begin{eqnarray}
\left\langle T_{\sigma _{1}}\right\rangle \left( t_{u}=1,\text{ }%
t_{d}=0\right) &=&-\frac{1}{2}, \\
\left\langle T_{\sigma _{2}}\right\rangle \left( t_{u}=1,\text{ }%
t_{d}=0\right) &=&0.
\end{eqnarray}%
Thus we get the scalar $g-$functions
\begin{eqnarray}
g_{1}\left( \theta \right) &=&\frac{-1}{-3+\left( 1+P\right) ^{3}/\left(
2P^{3/2}\right) -\left( \mathbf{m}_{A}\cdot \mathbf{m}_{B}\right) ^{2}},
\notag \\
&&  \label{Slonczewski} \\
g_{2}\left( \theta \right) &=&0,
\end{eqnarray}%
which show that only the Slonczewski torque exists in the ideal DSV. The
absence of any layer-thickness dependence in Eq. \eqref{Slonczewski} results
from the phase average across sufficiently thick normal metal layers. The
values of $g_{1}\left( \theta =0,\pi \right) $ are crucial to evaluate the
threshold current density $J_e^*$ needed for magnetization reversal of the
free layer since $J_e^*\propto 1/g_{1}\left( \theta =0,\pi \right) $. \cite%
{xrw3}

\begin{figure}[tbph]
\begin{center}
\includegraphics[width=8.0cm]{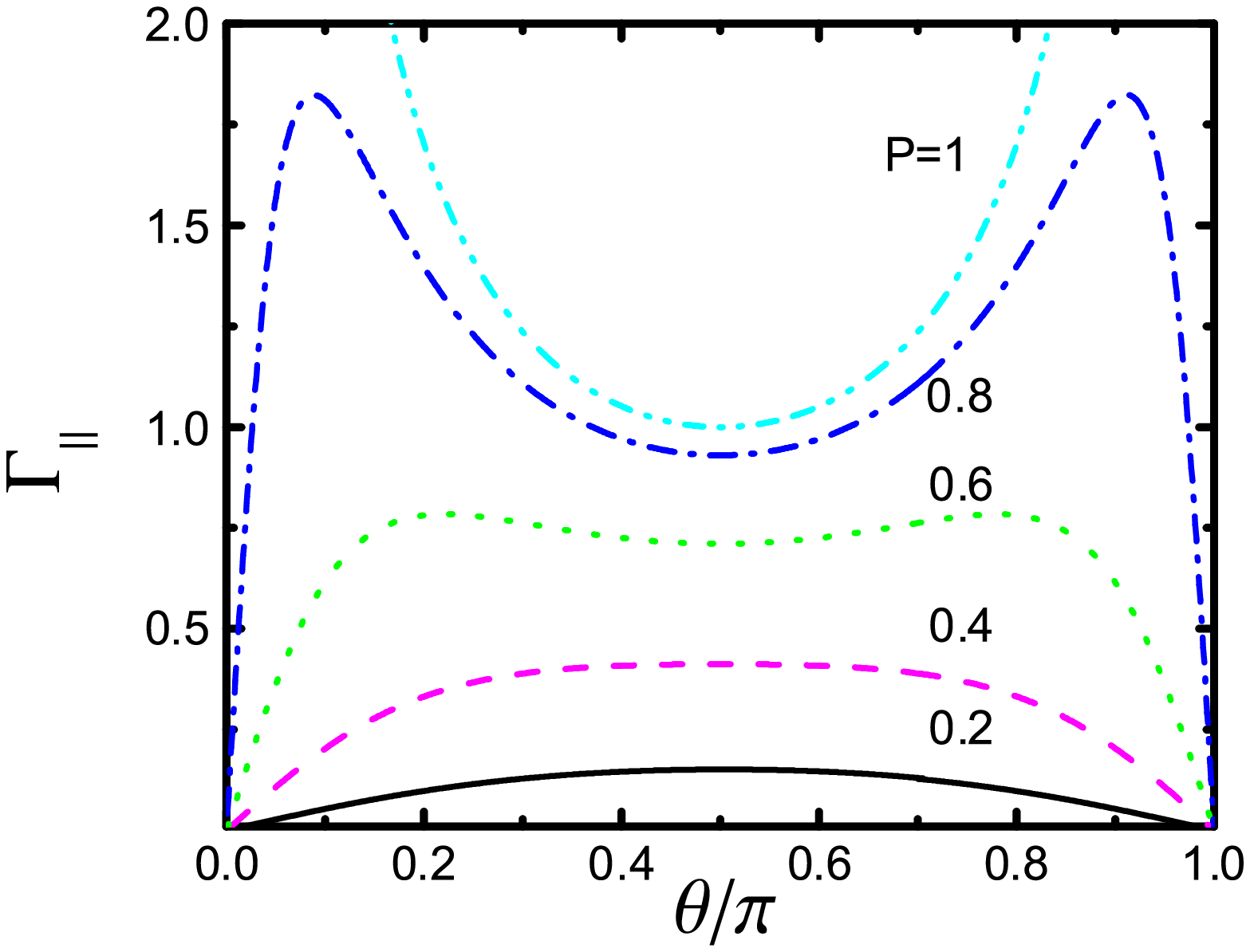}
\end{center}
\caption{(Color online) Slonczewski torque ${\Gamma }_{\parallel }$ versus
angle $\protect\theta $ for various polarization $P$ in the ideal case where
all FMs are perfect polarizers. The unit of torque is $\frac{\hslash }{e }%
J_{e}$.}
\end{figure}

Figure 4 is the magnitude of STT per unit current as a function of angle $%
\theta $ for various polarization $P$ in ideal DSV. The STT is symmetric
against $\pi /2$ due to the contributions from both fixed magnetic layers,
which is different from the result in conventional spin valve. \cite%
{Slon,Waintal,Krstajic} For polarization coefficient $P<1$ it generally
vanishes at $\theta =0$ and $\theta =\pi $ (shown in Fig. 4). However, if $%
P=1$, the torque is singular at $\theta =0$ and $\pi $. The divergence can
be understood mathematically and from a physics viewpoint. Mathematically,
the singularities at $\theta =0$ and $\pi $ are due to the factor of $%
g_{1}=-\left( \sin \theta \right) ^{-2}$ when $P=1.$ Physically this is the
consequence of perfect spin filter. In our model, every electron transfers
its angular momentum to local magnetization when it impinges the interface
of magnetic layer whose magnetization is not parallel to its spin. \cite%
{Slon} Meanwhile, the STT per unit current is defined as the spin transfer
per transmitted electron. \cite{Slon} However, in the case of $\theta =0$ or
$\theta =\pi ,$ perfect spin filter does not allow any electron transmitting
through the DSV, which leads to a zero electron transmission and results in
the divergence. The above argument can also be applied to the STT divergence
at $\theta =\pi $ in traditional spin valve in the original paper by
Slonczewski. \cite{Slon} Nevertheless, we find that the total STT $\mathbf{%
\Gamma =}-\frac{1}{2}\frac{\hslash }{e}J_{b}\mathbf{m}_{B}\times
\left( \mathbf{m}_{B}\times \mathbf{m}_{A}\right) $ if $P=1,$ does
not have such singularities.

Finally, we would like to compare the magnitude of STT in our DSV with that
obtained in traditional spin valve. \cite{Krstajic} In Ref. %
\onlinecite{Krstajic}, Krstaji\'{c} \emph{et al.} calculated the $g-$%
function in spin valve
\begin{equation}
g_{1}^{\ast }\left( \theta \right) =\frac{-1}{-4+\left( 1+P\right)
^{3}\left( 3+\mathbf{m}_{A}\cdot \mathbf{m}_{B}\right) /\left(
4P^{3/2}\right) },  \label{spin valve}
\end{equation}%
which is the same as the Slonczewski's result in Ref. \onlinecite{Slon}.

\begin{figure}[tbph]
\begin{center}
\includegraphics[width=8.0cm]{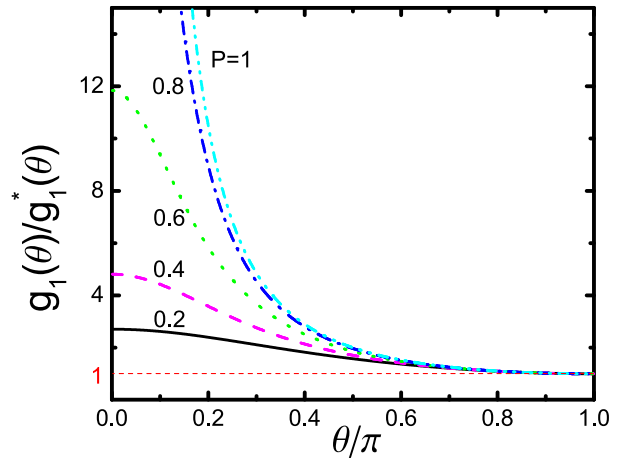}
\end{center}
\caption{(Color online) Spin transfer torque enhancement ratio $g_1\left(%
\protect\theta\right)/g_1^\ast\left( \protect\theta\right)$ versus $\protect%
\theta $ for various polarization $P$ in the ideal case where all FMs are
perfect polarizers. }
\end{figure}

We plot the ratio $g_{1}\left( \theta \right) /g_{1}^{\ast }\left(
\theta \right) $ as a function of angle $\theta $\ for various
polarization $P$ in Fig. 5. One can see that the STT is largely
enhanced when $\theta $ is acute, but approaches to that in usual
spin valve when $\theta >\pi /2$. Namely, the enhancement value
sensitively depends on both the orientation of free layer and the
polarization degree of electrons. Under small tilted angle $\theta $
and large polarization $P,$\ the enhancement ratio is dramatic,
which will substantially lower the required threshold current
density for magnetization switching. For instance,
$g_{1}/g_{1}^{\ast }=11.9$ if $\theta =\frac{\pi }{6}$ and $P=0.8.$
While the enhancement ratio approaches to $1$ (red dash line shown
in Fig. 5) when $\theta $\ is close to $\pi .$ The reason is that
there is no difference between a DSV and a usual spin valve when
layers FM$_{A}$ and FM$_{B}$ are antiparallelly aligned, since
FM$_{C}$ would not reflect electrons coming out of FM$_{B}$ in such
case. The results are qualitatively consistent with Fuchs \emph{et
al}.'s
experiment \cite{Fuchs} which shows that the reduction of threshold current switching FM$%
_{B}$ from parallel to antiparallel, with respect to FM$_{A}$, is
substantial, while it is only of modest size from antiparallel to
parallel in DSV compared to that in conventional spin valve.

From the qualitative physical picture and quantum mechanical calculations,
we conclude that STT can be greatly enhanced in DSV compared to that in spin
valve structure in the ballistic regime without spin accumulations. The
findings of the physics behind and the analytical formula of STT in this
emerging geometry should be interesting for both theoretical \cite{You,Balaz}
and experimental \cite{Fuchs,Watanabe} concerns. Micromagnetic simulation
based on our new $g-$function (Eq. \eqref{Slonczewski}) and
Landau-Lifshitz-Gilbert (LLG) equation \cite{Gilbert} will be a direction of
future research. The behavior of STT in asymmetric DSV when the widths of
two NMs are different, i.e., $l_{A}\neq l_{B},$ is also an interesting issue
for further investigation.

Although the advantage of STT enhancement is unambiguously demonstrated in
our results, the accuracy of the analytical formula still needs experimental
confirmation. We suggest to use a recently developed technique called
spin-transfer-driven ferromagnetic resonance (ST-FMR) \cite{Sankey} to
measure the angular dependence of the STT in DSV.

\section{SUMMARY}

In conclusion, we derive the STT acting on the free magnetic layer in a DSV
structure in the ballistic regime. A full-quantum mechanics description of
the STT is presented, which is valid for nanoscale DSVs where both the
electron mean-free path and the spin-flip relaxation length are larger than
the thickness of the spacers. \cite{Jedema} Using a quasi-one-dimensional
model and within the Slonczewski's approach, we obtained the analytical form
of the STT when all magnetic layers are perfect polarizers. Similar to the
results in the diffusive regime, the STT is dramatically enhanced in
comparison to that in a conventional spin valve although no spin
accumulation exists at the magnetic-nonmagnetic interfaces. Depending on the
orientation of free magnetic layer and the polarization degree of electrons,
the STT can be enhanced by a factor of a few tens. Our analytical $g-$%
function provides a theoretical base for the micromagnetic simulation of the
spin dynamics in DSV. The general cases when transmission and reflection
probabilities of free layer are different from zero or one are also
numerically calculated, which shows that the sign of the torque may change
under different transmission probabilities. These results should be useful
for the switching current reduction in magnetization reversal.

\section*{ACKNOWLEDGMENTS}

This work is supported by Hong Kong RGC grants (\#603007, 603508, 604109 and
HKU10/CRF/08- HKUST17/CRF/08). X.R.W. would like to acknowledge the
hospitality of Kavli Institute for Theoretical Physics China, CAS. Z.Z.S.
thanks the Alexander von Humboldt Foundation (Germany) for a grant.

\end{document}